# Wireless Sensor Networking for Rain-fed Farming Decision Support [*]

Jacques Panchard,
Panagiotis Papadimitratos,
Jean-Pierre Hubaux
LCA
EPFL, Switzerland
jacques.panchard@epfl.ch,
panos.papadimitratos@epfl.ch,
jean-pierre.hubaux@epfl.ch

P.R. Seshagiri Rao
Chennakeshava Trust
Karnataka, India
girickp@gmail.com

Madavalam S. Sheshshayee,
Sumanth Kumar
Univ. of Agriculture Sciences
Bangalore, India
msheshshayee@hotmail.com,
sumanth_ag@yahoo.com

## ABSTRACT

Wireless sensor networks (WSNs) can be a valuable decision-support tool for farmers. This motivated our deployment of a WSN system to support rain-fed agriculture in India. We defined promising use cases and resolved technical challenges throughout a two-year deployment of our *COMMON-Sense Net* system, which provided farmers with environment data. However, the direct use of this technology in the field did not foster the expected participation of the population. This made it difficult to develop the intended decision-support system. Based on this experience, we take the following position in this paper: currently, the deployment of WSN technology in developing regions is more likely to be effective if it targets scientists and technical personnel as users, rather than the farmers themselves. We base this claim on the lessons learned from the COMMON-Sense system deployment and the results of an extensive user experiment with agriculture scientists, which we describe in this paper.

## Categories and Subject Descriptors

C.2.m [**Ubiquitous Computing**]: Sensor Networks

## General Terms

Experimentation, Human Factors

## Keywords

Wireless Sensor Network, User Experiment, Developing Countries, Agriculture

## 1. INTRODUCTION

Today emerging countries face the antagonistic goals of developing their economies and improving their population's livelihood,

[*]This work was supported by the NCCR MICS and the Swiss Agency for Development and Cooperation

while sustaining their natural resources and environment. Environmental monitoring seems to hold great promise for tackling these issues. One typical example is air pollution monitoring, because this is a major concern in many urban areas of developing countries, especially in the mega cities of Asia. Water quality control is another opportunity: in developing countries, diarrhoeal diseases consecutive to the consumption of contaminated water are one of the main causes of child mortality.

Most of all, agriculture comes as a natural application, given the importance that climatic and physical parameters have throughout the development of the crop. Indeed, developing countries often have to cope with a large farming population who have seen their situation deteriorate in recent years due to the instability of market prices and the perceived effects of climate change.

As a typical example, the share of agriculture for employment in India is about 67%, with a majority of small land holdings. In Karnataka (Southern India), 87% of the farms have less than 4 ha and account for more than 50% of the total cultivated land, while 39% of the total are very small farms (less than 1 ha). These *resource-poor* farmers usually lack access to irrigation facilities and depend on rain-fed farming for their livelihood [1]. Their crop yields are highly *unreliable*, due to the variability in both the total rainfall and its distribution [2, 3]. Unlike industrial farming companies, they face daunting challenges, as is illustrated by the current wave of suicides throughout the country [4].

We performed a survey in 2004 to identify the user needs of the rural population in three small villages of Karnataka [5]: the themes of crop yield prediction, pest and disease control, and water levels in bore wells stood out prominently among farmers' responses. Practicing exclusively rain-fed farming, they would greatly benefit from a better knowledge of the field environment, and from more adapted strategies and practices.

To address these issues, it seems natural to assess the potential of wireless sensor networks (WSNs), in particular, the increased space and time resolution they bring to environmental data at an affordable price. WSNs are today viewed as an enabling technology for precision agriculture: irrigation management, frost prevention, nutrient or pesticides control are documented examples [6]. Still, it remained to be proven whether a similar knowledge could be used by the population who needs it most: the resource-poor farmers who practice rain-fed agriculture.

This has been the endeavor of the COMMON-Sense Net [8] project, which has investigated over three years the utilization of WSNs as a decision-support tool for rain-fed farming. Our goal was to help the farmers monitoring the physical farming environment, in order to understand more precisely the physical processes at hand, and to react optimally to changing conditions.

Our vision has been to bring the benefits of technology *directly* to farmers, in a participatory way. We deployed our system, described in Sec. 2, after identifying use cases with the locals. Over a long period of data collection and usage, we reached an impasse: numerous difficulties have emerged, essentially hindering our efforts to bring in a participatory manner the added value of WSN technology, through our system, to the farmers.

This has been primarily due to the farmers' alienation with the worlds of science and technology. Presenting the lessons learned through our three-year long effort and deployment, in Sec. 3, we are currently forced to consider as somewhat idealistic our initial objective. The resource-poor farmers could not really put enhanced environmental data in use effectively.

Based on these experiences, we moved on with a different approach: we investigate controlled-environment strategies related to rain-fed farming, such as developing new crop varieties or pest prediction measures. Accordingly, the *position* this paper takes is as follows: *Under the current conditions in developing regions, such as Karnataka, the targeted users for WSN-enabled applications, should be researchers, scientists, and technicians*. As our study case, long experience, and recent experiment we conducted in Bangalore, from November 2007 to February 2008, indicate, this orientation towards a new user group (of scientists) that can then advise or guide the farmers appears currently the only method to have an effective decision-support system for rain-fed farming.

This article brings this seemingly pessimistic yet realistic position to the attention of the community. Essentially, our concerted effort of deploying and running the COMMON-Sense Net system points that still scientists remain the preferred 'customers' of WSN technology. We present the learned lessons that led us to this shift away from our initial objective. Then, Sec. 4 describes the methodology for our user experiment to determine if our new user target group was an appropriate choice. In Sec. 5, we present our results, followed by a discussion, future work and conclusions.

## 2. THE COMMON-SENSE NET SYSTEM

### 2.1 Challenges and Opportunities

In order to assess the potential of environment monitoring for rain-fed agriculture and to draw technical constraints from the appropriate context, we decided to focus on the concern expressed for crop yield prediction. Semi-structured interviews with agronomists led to the following *use cases* [8]:

1. Deficit irrigation management: based on the water content of the soil, it is possible to design irrigation schemes that make use of limited water resources
2. Assessment of water conservation measures: using residual soil moisture, it is possible assess efficiency of measures such as building bunds, planting trees, mulching etc.
3. Existing crop models validation: using actual values of soil moisture, it is possible to evaluate crop models by comparing with predicted values

We opted for all these use cases to be developed in a *participatory manner* with the local farmers, in order to avoid the common pitfalls associated with the deployment of information systems in developing regions [9].

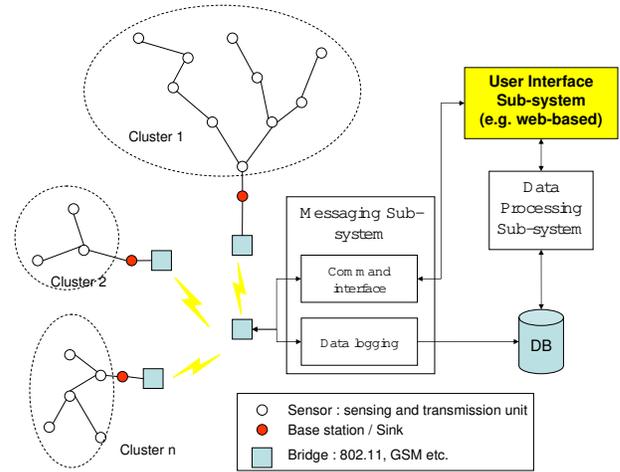

**Figure 1: COMMONSense Net system architecture**

### 2.2 System Architecture

The architecture of our WSN system is depicted in Fig. 1. So far, we have deployed two clusters for more than one year (two full cropping seasons) in a farming environment in the village of Chennakeshavapura (Karnataka) [8]. They measure and provide a detailed picture of the soil moisture conditions in 10 homogenous patches, all rain-fed, with different crop varieties and different soil characteristics. In addition, one node records air temperature, humidity and atmospheric pressure over the same period. We detail and justify hereunder the system's main design characteristics:

**Reduced and Targeted Data Types:** The main focus being the water available in the soil for the crop, we designed a simplified data acquisition board allowing the instrumentation of two soil moisture probes of the ECH2O [10] type, so that the top and bottom of the plant's root zone could be monitored.

**Power Management:** We considered the use of solar panels but ruled it out because of the exposure of such hardware to theft and tampering. Whereas a wireless sensor in a box can be camouflaged in an outdoor field, a solar panel has to stand prominently and is easily identified as an object of value. We turned instead to an original MAC plus routing scheme, Dozer [11]. Our most recent tests indicate a lifetime of more than 7 months (more than one cropping season) for nodes deployed with this cross-layer protocol in the conditions of our experiment.

**Periodical Data Collection:** Because we were interested in the continuous evolution of phenomena (in particular, water content in the soil), we chose a periodic data collection model that allowed for the logging of extended data sets in a centralized database for later retrieval and analysis. Because of the uncertainty regarding the necessary time resolution, the data sampling period was set to two minutes. This value is higher than needed a priori by any agricultural application, but as it was compatible with the expected lifetime of the network, it was chosen as a benchmark from which to assess the ideal data rate a posteriori.

**Sparse Network:** 2-3 sensors were needed per homogenous soil parcel, which is in the order of one ha. If one is to avoid deploying extra nodes for the sole purpose of relaying traffic, a radio with communication range above the average performance of wireless sensors is to be used. We finally opted for the tinynode [12] platform, which allowed us to obtain a communication range of 200m and above in a consistent manner.

**Cluster-based Hybrid Architecture:** The interconnection of far-apart clusters with the centralized server necessitated the design of a hybrid architecture with the cluster heads implementing a bridge to the server. In the absence of cellular phone coverage, we implemented in a first phase a IEEE 802.11 bridge. As GSM coverage improved in 2007, we migrated to a GPRS-based bridge that removes the necessity to maintain a local server at the deployment site.

**Web-based Interface for Data Retrieval:** We developed a web-based application for the retrieval of cluster-based information (such as maps and per node information) as well as individual sensors data display and download. This application can be accessed at: http://csn.epfl.ch. Basic instructions on how to access these data can be found in Appendix.

## 3. LESSONS LEARNED

Deploying a sparse network in a remote and uncontrolled environment raised critical deployment issues. Fluctuations of the radio channel caused by the growth of vegetation during the cropping season had a severe impact on the network connectivity. To diagnose such problems and other software or hardware failures would require a constant presence of communication engineers on the deployment site, which is difficult to achieve in a rural setting.

The main lesson to be learned from this deployment, however, was the difficulty of implementing and testing convincingly our use cases in the field. In order to achieve this, the collaboration of the farmers was required to protect the hardware, to report regularly on field conditions and to give feedback on the added value brought by the technology. Unfortunately, we observed an initial distrust of the population towards the technology and the presence of scientists in the field.

Informal discussions with local stake-holders indicated that the population of small farmers has an experience of being systematically left behind in the innovation process. As a consequence, they feel an general distrust for scientists and technologists, who are perceived as wanting to take advantage of them for their own benefit rather than helping them. As a consequence, we had some experience of theft and tampering with the hardware deployed in the fields. This combined with the harsh environment (heavy rains, intrusion of wild-life) resulted in the loss of more than ten wireless sensors over one year. Given the moderate size of our deployment, this represents a significant phenomenon. A second obstacle was the difficulty to translate the scientific terminology of environmental science (soil moisture, evapotranspiration, etc.) into the language of the field. Farmers think rationally about their business, but mathematic concepts such as probabilities or percentage (of risk, of soil moisture, etc.) are usually not part of their language. Since a decision support system can only inform its users about alternatives, and not make decisions for them, there is a non-trivial language barrier to overcome. Finally, the current uncertainty about the benefit to cost ratio of the technology did not encourage active collaboration.

Both the difficulty to trace the technical problems and the impossibility to create a partnership with farmers represented serious obstacles. This called for a change of paradigm: in our experience, the promise of ubiquitous computing will have to wait for maturation of both the technology and the users before being fulfilled. Instead, managing the technology in a controlled environment with the participation of committed users can lead to rapid results, provided we can ensure a spill-over effect on the farming population. In the next section, we explain why we decided to focus on scientists working on applied research for rain-fed agriculture, and how we verified the appropriateness of this new approach.

In order to tackle the deployment issues caused by the difficulty to handle wireless sensors, we also developed an innovative deployment and maintenance support tool based on audio (rather than visual) feedback. This tool is supposed to help positioning the nodes and monitoring their health status without the need of a heavy infrastructure. It is described in details in

## 4. CHARTING THE PARADIGM SHIFT

The path from user needs to precise specifications of a system is not an easy one to trod. In the previous section, we identified a strong necessity to find a mediator between the technology and the target population. Agricultural scientists are ideally placed to define use cases. However, this is no trivial matter, because for them sensor data represents a new and unfamiliar context. Most of the scientists we interacted with are not familiar with sensor data at high resolution in time and space provided by a large number of data gathering points with uniform accuracy.

### 4.1 Choosing the Target Population

For these two reasons, we decided to set-up an experiment where we confront soil physicists, crop physiologists, entomologists, pathologists and agronomists with the results gathered from the field by our deployed prototypes.

The reason for selecting such a various user basis is double. On one hand, we wanted to find the largest scope of use for the WSN technology in the context of rain-fed farming, and did not want to restrict ourselves to our own preconceptions. On the other hand, as different disciplines can have various data requirements, it was important to know whether an appropriate system design could meet all of them.

There are several types of institutions where such professionals are likely to work, each of them with its own goals and agenda:

1. Academia: scientists doing research in agricultural departments. For them, two competing goals are at stake. Doing research that provides them with scientific impact and visibility, as well as solving practical problems.

2. Government: scientists working either as advisors for policy makers or as implementers of programs at the local level. Marginal farming is only one aspect of their concern, which is agriculture as a socioeconomic sector.

3. Non-governmental agencies: NGOs focusing on rural development often are innovative in terms of agricultural practices. Thus, they are regularly interested in applied research.

We did not extend our survey to the industry of agriculture inputs or to corporate agriculture, although these two sectors are likely users of the wireless sensor networking technology. In the first case, we did not want to get involved in the controversy surrounding the effects of large seeds providers on the livelihood of small and marginal farmers in India. In the second case, the type of agriculture practiced (mechanized farming, on-demand irrigation, precision agriculture with high added-value) was considered too different from our focus of interest, namely rain-fed farming.

### 4.2 Goal and methodology

We interviewed 30 people from the backgrounds detailed above, following both a qualitative and a quantitative approach. The goal of the experiment was to assess the use that agriculture scientists would make of the data that are collected by the COMMON-Sense Net system.

The experiment was scheduled to run for 2 weeks in November 2007. We asked the scientists a series of questions about the value of environmental data for them.

The goal of the experiment was to understand precisely:

1. What are the types of environmental data they can make use of, and how?
2. What is the spatial diversity they will use for the data?
3. What is the time granularity they will make use of in their task?

In order to answer to these questions, we used three complementary approaches:

*Structured interviews:* As a preliminary, some general questions were asked to the participants in a *general questionnaire* before they tested the interface during a two-weeks long study. These questions served to assess their current view of the field. Participants were asked to answer more concrete questions in a *detailed questionnaire* during the experiment. A detailed description of the questionnaires can be found in [13].

*Behavior observation:* It is risky in such an experiment to rely solely on users' opinions. This is why the scientists were encouraged to provide data sets in order to substantiate their answers (in the form of graphs or numbers). Moreover, the queries they made to the database in order to retrieve data were recorded. In particular, the time and space granularity of the data that they consulted was logged into a database. This makes it possible to analyze the scientists' behavior as well as their discourse.

*Semi-structured meetings:* Finally, a debriefing took place after the experiment, in order to allow the participants to share their impressions in an informal discussion. As it will appear clearly in the next section, this part turned out to be surprisingly rich in information.

## 5. EXPERIMENT RESULTS

### 5.1 Questionnaires

An extensive list of the data and their interpretation can be found in [13]. All participants but one identified environmental information as an important input for the study of rain-fed farming. As for parameters, soil water content and temperature are considered the most important to monitor, as shown in Fig. 2.

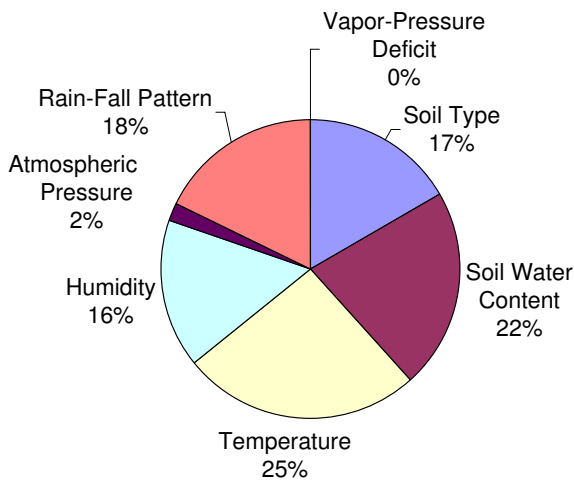

**Figure 2: Critical parameters as identified by the experiment's participants**

The participants also had to characterize the parameters they chose with regard to the required spatial density, sampling frequency, and tolerated error. Regarding spatial density, the parameters can be classified into two categories: the parameters considered with a low spatial variability, i.e. one kilometer and above, such as temperature, rain fall and atmospheric pressure. And the parameters demanding high spatial variability (from 500m down to one hectare): only soil moisture belongs to this category. Interestingly, soil type shows a bimodal result, about half the users considering that a single measurement point is enough, and the other half considering that it should be performed at least every few hundred meters.

The required measurement frequency shows a wider distribution. It is interesting to note, however, that in all cases but one (soil moisture), the lower limit is the day. Only for soil moisture were hourly or more frequent measurements deemed appropriate, and only for a minority of users (less than 10 percent).

As for the error tolerance for each parameter, the participants tend to require high precision (less than five percent error).

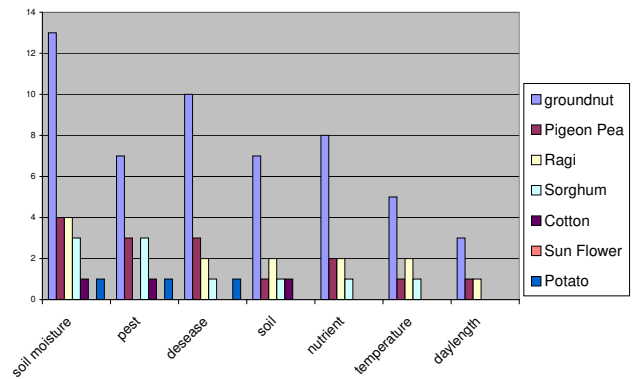

**Figure 3: Detailed Questionnaire: Constraints per crop type (expressed in number of answers)**

Fig. 3 shows the parameter distribution per crop. This table confirms that for each crop the parameters to take into account are the same, except pest, which is not considered to be an issue for ragi.

The results about parameters, their variability and error tolerance are consistent between the general questionnaire and the detailed questionnaire. In other terms, users did not noticeably change their mind after using the web application. The next section clearly explains why.

### 5.2 User Activity Logging

This is the set of metadata that was generated by the logging of participants' interactions with the web application. This part of the experiment led to inconclusive results. Out of the 30 participants, only six actually used the on-line application at some point. All of them were PhD students and post-docs. No senior researcher spontaneously used the on-line application.

The participants who used the application did an average of 3 queries to the system, mostly to look at the soil moisture status.

This paradoxical disinterest for the on-line application is discussed in the next question. It made the debriefing meetings very important, in order to understand the mismatch between the interest manifested in the survey and the actual use of the application.

## 5.3 Debriefing Meetings

Initially, the debriefing meetings were intended to gather the opinions of the participants in a more informal manner than during the experiment. However, in light of the mismatch mentioned in the previous section, they became a crucial element of the experiment. The goal was to find out why the users had not used the application as expected, and to assess their real level of interest.

Instead of asking these questions directly, which would have been likely to bias the answers, we chose to address concrete use cases. If people were able to come up with original use cases, that meant they had conducted a reflection about the tool. Moreover, we could then talk concretely about upcoming partnerships, an extra-measure of their interest, and a critical conditions for the continuation of the project.

The results were encouraging. We found four compelling use cases, and in each case a concrete interest in using the technology provided. Precise details about the requirements of each use case can be found in [13].

### 5.3.1 Soil Science

Provided we can adapt nutrient sensors to the wireless nodes, there is research to be conducted in the response at the root zone to different strategies of nutrient application and irrigation. The main objective would be to observe the variation in nitrogen, phosphorus and potassium content, in the context of nutrient dynamics under a system of multiple crops and trees.

Soil salinity, pH and moisture are available today. Monitoring of specific nutrients, however, remains an open question.

### 5.3.2 Entomology

The observation of pests present in the crop field shows that their activity depends on the weather, especially rain fall, soil moisture and soil temperature. There is a clear correlation between the rain patterns and the emergence of adults of the insects from the soil.

The hypothesis to verify is whether the insect's activity depends on soil moisture evolution and accumulation of soil heat in the weeks prior to emergence. If the soil moisture conditions are not favorable to them during pupation, a large percentage of the population might die.

Soil moisture, temperature sensors in specific regions of endemic populations of these pests (sampling various soil texture typologies) will help to investigate, understand their biology. This would make it possible to provide advance information on the intensity of pest damage to farmers.

### 5.3.3 Crop Physiology

This use case is about the precise assessment of the ratio between the water that is transpired by the plant and the water that is evaporated, in other terms the plant's water efficiency. The possibility to achieve crop improvement through selection would have a positive impact on yields achieved in rain-fed farming. For this, it is necessary to test plants with different genotypes obtained by cross-breeding and to assess which one has the best ratio of biomass production per water used.

The method used for this test today is gravimetric method. For this, plants in pots are used. The pots are filled daily with water up to field capacity. The next day, they are weighted to assess the water lost in evapo-transpiration. Bare plots are used as a benchmark to assess the effect of pure evaporation.

The goal is to replace the gravimetric method with soil moisture sensors that would give directly the volumetric content of water of the soil. The tedious weighting procedure could then be avoided.

### 5.3.4 Water management

For a large NGO conducting applied research in the area of rain-fed farming, wireless sensor networks are perceived as a promising validation tool. Two experiments are envisioned:

1. the possibility to increase soil water-retention capacity through different measures, such as fertilizer, mulching, etc.

2. assessing the efficiency of underground drip irrigation. Here, the goal is to bring the water directly to the root zone of the plant.

For both experiments, soil moisture is the ultimate measure of success or failure.

## 6. DISCUSSION

Potential users expressed keen interest in several cases. In particular, a major NGO working in the field of dry-land farming throughout India expressed interest in the WSN technology. Such promising results must be tempered by the low response obtained by the application use, which we address in section 6.3.

## 6.1 Usefulness

From the questionnaires, there is a large consensus as to the utility of using finer-grained environmental data for rain-fed agriculture.

The level of detail, at which scientists answered the on-line survey indicates a high level of interest and curiosity on their part. Such an interest was already perceptible at the inception of the project. However, the creation of precise use case was not possible then.

This gap was filled during the individual interviews with scientists coming from academia, as well as the non-governmental sector. With four precise use cases and potential partners clearly identified, the initiative is now in the hands of the information and communication systems specialists.

One central question is the potential of information-sharing with farmers. Will the results ever leave the lab and scientific reports to materialize in the field? According to our interviews with government officials [13], the Indian institutional framework is very clear: The agriculture scientists are expected to provide scientific evidence of phenomena, to investigate preventive or corrective actions when appropriate, and to publish recommendations that are used by the agriculture department to relay information to the public (which is referred to as extension work). The case of the non-governmental sector is different. Large NGOs are conducting applied research aimed at improving farming practices in dry-land management. In this case, the scientists work directly in contact with the farmer. This makes them privileged partners.

## 6.2 Usability

New types of sensors were mentioned in the course of the experiment. Most prominently, in-situ chemical sensors that could sense the concentration of nutrients (nitrogen, phosphorus, potassium) in the soil are mentioned repeatedly. The development of such sensors is still at an experimental stage, but some recent advances have been made for low-cost sensors using ion-selective electrodes [14].

The accuracy is a recurring concern. However, the exact precision to which sensors need to operate is still an open question. In general, it is more important to be roughly right than precisely wrong.

Parameters linked with the soil are pinpointed as having a high space variability. As a consequence, any high-precision technology that comes at a price such that it is not possible to diversify the readings is not going to be usable. With low cost sensors, a relatively high error can be compensated by spatial diversity, which allows for the statistical elimination of outlying measurements.

The sampling period of the data (time variability) is still the object of uncertainty. Measures shorter than a day are not a priori taken into consideration by scientists in the framework of an applied research for agriculture. At present, the researchers mostly want daily data, since they use data of similar granularity obtained from conventional measurements without sensors. During debriefing sessions, however, we gathered evidence that the responses are likely to change with time spent on reflection and experience with high-resolution data. Indeed, when prompted by one of the authors with background in agriculture about possible uses for research, the participants acknowledge in a majority of cases that such data could be used in the framework of their research.

It appears that certain elements of the current responses, particularly those related to time resolution, will change after some experience and / or contemplation on use of high time resolution data. Researchers have always viewed data gathering as a major constraint in research design and conceptualization. The current experiment presents a completely contrasting situation with the provision of very rich data in both time and space for parameters of interest. It is in this light that we suggest a co-learning process for agriculture researchers and sensor technology providers to evolve better and meaningful use cases.

### 6.3 Use

We tried to provide possible explanations about the paradoxical low level of use recorded during the experiment. It cannot be ruled out that this reflects the actual disinterest of the participants. However, that would be contradictory with all the other results of the survey.

Moderate computer literacy is a possible explanation. A similar result was obtained during a previous experiment conducted in 2006, specifically about interface design [15]. At that time, we had seen a clear difference of behavior between scientists used to working with environmental sensors and computers, and those who did not have this expertise. As a consequence, we sought to improve the interface, but such a gap remains, as illustrated by the difference in handling the application from senior scientists and their younger, computer-savvy students.

However, this is only a partial explanation to the observed phenomenon. As a matter of fact, despite their computer literacy the students and post-docs did not extensively use the application. A posteriori, this can be explained by the difficulty in finding a one-size-fits-all scenario for the participants of the survey. We had not realized how diverse the concerns of agricultural scientists could be. It is likely that the questions asked made little sense to most of the participants to the survey.

## 7. CONCLUSION AND FUTURE WORK

We performed an experiment about the usefulness and usability of wireless sensor networks for rain-fed agriculture. The usefulness to provide finer-grained environmental data for applied research in this field was clearly identified. The technology gap remains, however, and can only be bridged with a highly dynamic process of interaction with potential users. Building appropriate partnerships and exploring the outcomes of medium and long-term deployments of WSNs is the next step. The interest from the non-governmental sector is the most promising because it presents the most direct contact with the intended beneficiaries of this technology: the small land-holding farmers.

## 8. REFERENCES

[1] H. Diwikara and M. Chandrakanth, "Beating negative externality through groundwater recharge in India: a resource economic analysis," *Environment and Development Economics, Cambridge University Press*, vol. 12, pp. 271–296, 2007.

[2] S. Gadgil, Y. P. Abrol, and P. Rao, "On growth and Fluctuation of Indian Foodgrain Production," *Current Science*, 1999.

[3] I. Abrol, "India's agriculture scenario," *Climate variability and agriculture*, pp. 19–25, 1996.

[4] S. Mishra, "Suicide of Farmers in Maharashtra," *Indira Gandhi Institute of Development Research*, 2006. [Online]. Available: citeseer.nj.nec.com/557578.html

[5] S. Rao, M. Gadgil, R. Krishnapura, A. Krishna, M. Gangadhar, and S. Gadgil, "Information Needs for Farming and Livestock Management in Semi-arid Tracts of Southern India," CAOS, IISc, Tech. Rep. AS 2, September 2004.

[6] T. Wark, P. Corke, P. Sikka, L. Klingbeil, Y. Guo, C. Crossman, P. Valencia, D. Swain, and G. Bishop-Hurley, "Transforming agriculture through pervasive wireless sensor networks," *IEEE Pervasive Computing*, vol. 6, no. 2, pp. 50–57, 2007.

[7] J. Panchard, S. Rao, T. V. Prabhakar, J.-P. Hubaux, and H. S. Jamadagni, "Common-sense net: A wireless sensor network for resource-poor agriculture in the semiarid areas of developing countries," *Information Technologies and International Development*, vol. 4, no. 1, pp. 51–67, 2007.

[8] J. Panchard, E. Costanza, J. Freudiger, A. Ouadjaout, B. Bostanipour and J.-P. Hubaux, "Making the Invisible Audible: Acoustic Interfaces for the Management of Wireless Sensor Networks," submitted at *The 6th ACM Conference on Embedded Networked Sensor Systems (Sensys)*, Raleigh, 2008.
Panchard, Jacques ; Costanza, Enrico ; Freudiger, Julien ; Ouadjaout, Abdelraouf ; Bostanipour, Behnaz ; Hubaux, Jean-Pierre

[9] R. Heeks, "Information Systems and Developing Countries: Failure, Success and Local Improvisations," *The Information Society*, 2001.

[10] "ECH2O Soil Moisture and Microclimate Monitoring," documentation available at http://www.ech2o.com/.

[11] N. Burri, P. von Rickenbach, and R. Wattenhofer, "Dozer: ultra-low power data gathering in sensor networks." in *IPSN*, T. F. Abdelzaher, L. J. Guibas, and M. Welsh, Eds. ACM, 2007, pp. 450–459.

[12] "tinynode wireless sensor from Shockfish," http://www.tinynode.com.

[13] J. Panchard, S. Rao, and S. Sheshshayee, "Wireless sensor networks for applied research on rain-fed farming in India: an exploratory user experiment," Tech. Rep., 2008, available at: http://people.epfl.ch/jacques.panchard.

[14] H. Kim, J. Hummel, K. Sudduthb, and P. Motavalli, "Simultaneous Analysis of Soil Macronutrients Using Ion-Selective Electrodes," Ph.D. dissertation, 2007.

[15] J. Panchard, "Computer-assisted Cognition: Using Wireless Sensor Networks to Assist the Monitoring of Agricultural Fields," EPFL, Lausanne, Tech. Rep., 2006, available at http://people.epfl.ch/jacques.panchard.

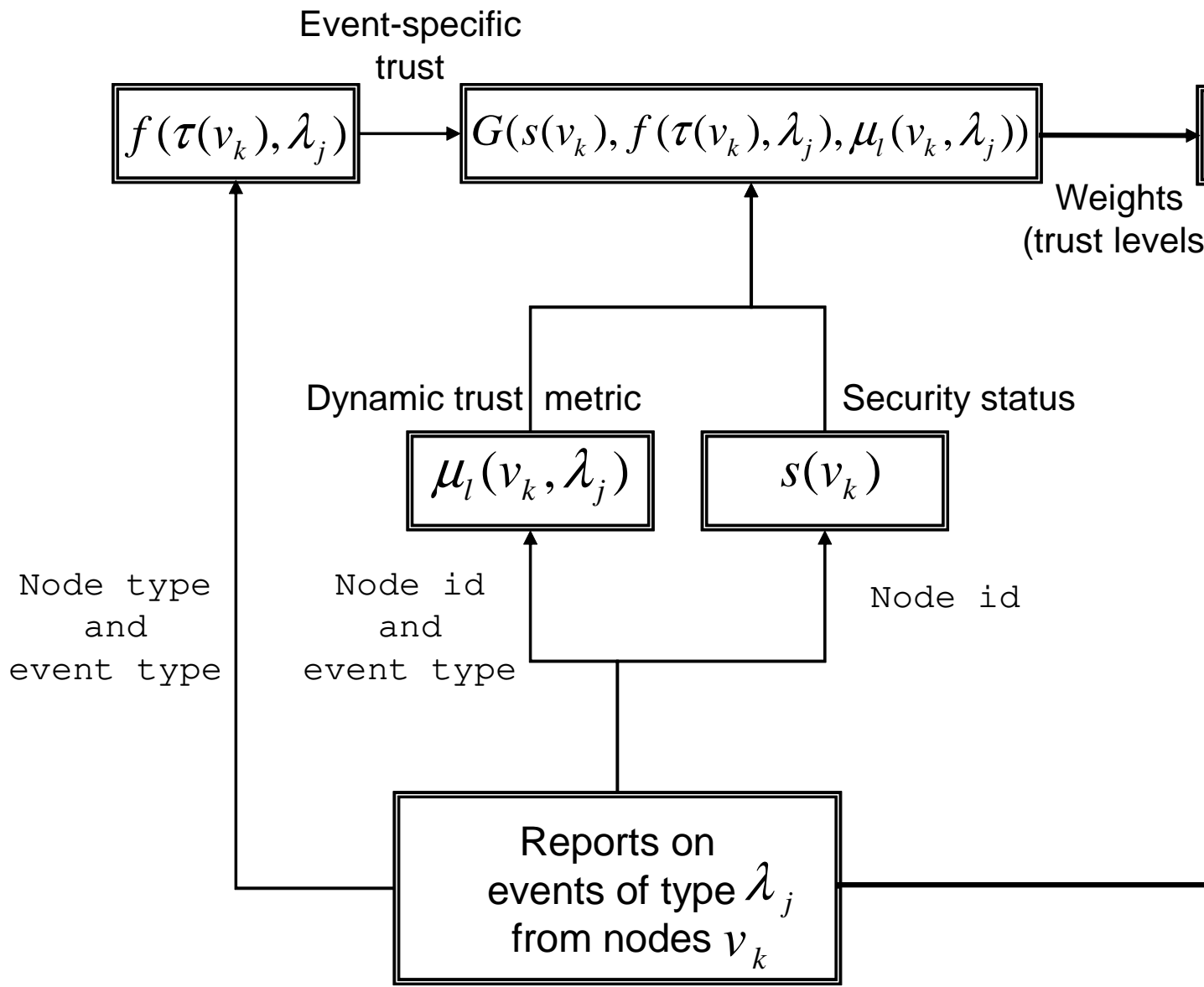